# UXR PoV for Neuroinclusive Emotion Regulation

## Developing a UXR Point of View for Neuroinclusive Emotion Regulation with Generative AI

Melike Akca
School of Computing and Engineering, Bournemouth University Poole, UK, makca@bournemouth.ac.uk

Mona Giff
London, UK, monaegiff@gmail.com

Deniz Cetinkaya.
School of Computing and Engineering, Bournemouth University Poole, UK, dcetinkaya@bournemouth.ac.uk

Huseyin Dogan.
School of Computing and Engineering, Bournemouth University Poole, UK, hdogan@bournemouth.ac.uk

Stephen Giff
Google Redmond, Washington, USA, sgiff@google.com

Attention-deficit/hyperactivity disorder (ADHD) is a psychiatric disorder which presents itself in individuals through patterns of developmentally inappropriate levels of inattentiveness, hyperactivity, and impulsivity, with difficulties in decision making and emotional regulation (ER). ER difficulties constitute a significant yet insufficiently addressed component of ADHD, contributing to emotional instability, reduced quality of life, and increased vulnerability to comorbid mental health conditions. Although digital and AI-based interventions have expanded access to ER support, many existing systems remain limited by weak theoretical integration, insufficient accommodation of neurodiversity, and a lack of structured user experience research (UXR) methodologies, that bridge psychological insight with design practice. This paper introduces a Generative AI–augmented UXR methodology, grounded in the UXR Point of View (PoV) Playbook, to support the design of emotionally intelligent and Neuroinclusive digital ER interventions for adults with ADHD. The approach integrates empirical evidence with established psychological frameworks Dialectical Behaviour Therapy (DBT)**,** Self-Determination Theory (SDT), and the COM-B behavioural model and leverages Generative AI as a co-analytic tool to support synthesis, hypothesis formation, and design articulation. The methodology is operationalized through a four-stage UXR process encompassing AI-supported hypothesis generation, foundational planning, insight generation via Building Blocks, and the construction of stakeholder-specific PoV narratives. This process results in a set of ten theory informed UXR Play Cards that translate psychological mechanisms and empirical findings into actionable design guidance. The primary contribution of this work is a replicable, bias-aware framework for integrating Generative AI into UXR practice, advancing human-centred and Neuroinclusive approaches to digital mental health design.

**ACM Reference Format:**

## UXR PoV for Neuroinclusive Emotion Regulation

### 1 INTRODUCTION

Attention-deficit/hyperactivity disorder (ADHD) is a psychiatric disorder which presents itself in individuals through patterns of developmentally inappropriate levels of inattentiveness, hyperactivity, and impulsivity, with difficulties in decision making and emotional regulation (ER) [1]. ER has emerged as a critical determinant of mental health, adaptive functioning, and overall wellbeing in individuals with ADHD [1]. ER difficulties result in challenges such as impulsivity, affective instability, and frustration intolerance [2]. These difficulties frequently disrupt social and occupational functioning and intensify core ADHD symptoms, contributing to a cascade of secondary psychopathologies, including anxiety and depression, which in turn elevate risks of social exclusion and long-term health inequities [3]. However, despite growing advances in the neuropsychological understanding of ADHD, challenges related to ER remain persistent and are often under-addressed in [4].

To address these challenges, digital health interventions (DHI) have emerged as accessible tools for supporting ER. A DHI is a discrete functionality of a digital technology which is applied to achieve health objectives [5]. DHIs aim to make mental health interventions more accessible, as they are 'ecologically' embedded, by using a smartphone-delivered method [5]. As such, they are a promising method for improving access to care and enhancing user engagement. However, their effectiveness for ADHD populations remains limited, largely due to the absence of design methodologies which are both theoretically grounded and response to the cognitive and emotional variability inherent in ADHD [6] These methodologies often use User-Centred Designs, but while it offers a valuable, empathy-driven framework emphasising participatory approaches and iterative refinement it often lakes the theoretical rigor needed to translate psychological models of ADHD into concrete, testable design outcomes [4].

The psychological models which can be used to address emotional dysregulation in ADHD are three complementary models. The first is Dialectical Behaviour Therapy (DBT), which provides a clinically validated structure for emotional scaffolding and mindfulness-oriented feedback, important for improving ER [7]. The second is Self-determination Theory (SDT). This model emphasises intrinsic motivation and autonomy as critical drivers of engagement and persistence [8]. Finally, the COM-B model establishes a systems view of behavioural (b) change linking capability (c), opportunity (o), and motivation (m) [9]. Together, these models can be used to inform a DHI for ADHD and ER. However, as mentioned previously, a current major weakness of DHIs is that they rely on generalised behavioural paradigms which insufficiently address neurodivergent traits such as sensory sensitivity a fluctuating motivation [6].

The User-Experience Research (UXR) Point-of-view (PoV) framework, as suggested by Dogan et al., aims to formalize the UXR process which synthesizes insights from UXR into actionable design decisions [10]. This involves integrating empirical evidence, psychological theory, and stakeholder priorities. This framework is based on the theory of Situational Awareness (SA) as proposed by Endsley, in which SA is defined by the perception of elements in the environment within a volume of time and space, the comprehension of their meaning, and the projection of their status soon [11]. SA is to the concept of a cognitive hierarchy in which information is said to come from correlated data; information is then converted into SA and becomes knowledge, and knowledge is used to predict the consequences of actions, which in turn, leads to understanding [12]. Dogan et al. argue that their PoV framework correlates with this cognitive hierarchy, in which the final step is wisdom, or in other words, the generated UXR PoV. It is a pyramid-shaped framework in which the bottom tier is (1) foundation, then (2) data collection, the third is (3) insight generation, and the top of the pyramid is (4) a PoV [10, 25]. The tangible result of this process is to generate 'play cards' which include a potential obstacle for progressing a PoV and best practices associated with this. These cards can then be used to craft PoV and stakeholder narratives. This essentially formalizes UXR into a comprehensive research-to-design translation pipeline which moves from raw data to actionable insight, and from insight to design decisions that are empirically testable, clearly communicable, and aligned with both user needs and stakeholder priorities. As such, this may be used to address the issue of weak methodologies in the design of DHIs [10, 25].

Furthermore, AI can be leveraged to produce insights from UXR using this UXR PoV framework. Giff et al. suggest four stages to develop an AI-Powered UXR PoV [13]:

1. Leverage GenAI and the UXR PoV Framework
2. Establish a Foundation Plan/Roadmap
3. Insight Generation and Card Development
4. Craft PoV Narratives for Stakeholders

Emerging work on AI-supported UX research suggests that Generative AI can facilitate progression through structured research frameworks by supporting data synthesis, hypothesis generation, and insight articulation. Such capabilities align with recent studies highlighting GenAI's role in augmenting UX practitioner workflows and enabling AI-assisted POV construction [13].

Therefore, the current paper aims to adapt the UXR PoV framework [10] for digital ER in ADHD, using GenAI as a cognitive partner to analyse evidence and simulate stakeholder reasoning, while integrating the three complimentary frameworks (DBT, SDT, and COM-B). This paper is concerned with how GenAI can translate research outcomes into actionable UXR PoV and narratives. Furthermore, it aims to address existing gaps in emotional intelligence, neurodiversity inclusion, and AI-UX co-creation, through a theory-grounded, user-centred design methodology.

## 2 METHODS

### 2.1 Study Design Overview

This study employed an AI-augmented systematic literature review (SLR) combined with User Experience Research (UXR) synthesis using the UXR Point-of-View (PoV) framework [10]. The method integrates PRISMA 2020 evidence synthesis with Generative AI (GenAI)–assisted reasoning to translate emotion regulation (ER) research in adults with ADHD into actionable UX PoVs for digital health interventions (DHIs). The methodological process followed the four-stage structure of the *Developing an AI-Powered UX Research Point of View (PoV)* workshop [13]: (1) leveraging GenAI and the UXR PoV framework; (2) establishing a foundational plan and roadmap; (3) insight generation and UXR Play Card development; and (4) PoV narrative construction and stakeholder communication.

### 2.2 Sample and Literature Identification

A systematic literature search was conducted across ACM Digital Library, PubMed, and Scopus, covering publications from 2020–2025. Search terms combined *ADHD*, *emotion regulation*, and *digital interventions* (e.g., mobile apps, web-based tools, AI-assisted systems, VR). Studies were screened according to predefined inclusion and exclusion criteria. Following PRISMA 2020 procedures, 42 empirical studies were included in the final dataset. All included studies were coded in Microsoft Excel using a PRISMA-aligned structure, including bibliographic data, population characteristics, intervention type, outcome measures, and ER-relevant findings. A separate worksheet documented themes identified through qualitative synthesis. To support AI-assisted analysis, the following materials were uploaded to each GenAI system:

1. the PRISMA-organized Excel dataset,
2. the UXR PoV framework papers (Giff et al., 2025; Dogan et al., 2024) [10, 25], and
3. the UXR previously published papers and workshop PowerPoint slide,
4. the UXR PoV Playbook website and example Play Cards.

A systematic literature review was conducted on the full dataset, followed by a thematic analysis using Braun and Clarke's (2006) six-phase method [14]. Identified themes informed both human-led interpretation and AI-assisted synthesis. Comparative observations across AI systems were analysed descriptively to assess differences in interpretive depth, theoretical integration, and UX relevance. Four Generative AI systems were evaluated: ChatGPT (v5.2)**,** NotebookLM, Gemini (v3)**,** and Microsoft Copilot**.** Interactions with ChatGPT and Microsoft Copilot were conducted by the first author, while Gemini and NotebookLM were conducted by the second author**.** All systems were prompted using the same predefined prompt set, submitted sequentially within a consistent computing and browser environment, to ensure procedural consistency, traceability, and comparability across models.

### 2.3 Procedure and Prompts

#### Stage 1: Leveraging GenAI and the UXR PoV Framework

To extract design-relevant themes and formulate 10 testable hypotheses from systematic literature review data using GenAI-assisted synthesis. Using an AI-augmented systematic literature review (SLR) approach, researchers synthesized PRISMA-structured data on emotion regulation (ER) and ADHD from an uploaded Excel dataset.

Generative AI was employed to identify thematic patterns, cluster key intervention features, and summarize common engagement challenges. Based on this analysis, hypotheses were generated linking user experience (UX) variables such as feedback, autonomy, and visual predictability to ER outcomes. Prompts for this stage:

- *"Analyse this PRISMA_ADHD_ER Excel dataset and summarise recurring design–emotion–behaviour patterns."*
- *"Cluster themes around attention, feedback, motivation, and emotion regulation success."*
- *"Generate one hypothesis per theme linking design features to emotional regulation, relying on the UXR PoV document, the UXR PoV website, and the published papers."*

While Stage 1 identifies empirically grounded themes and generates testable hypotheses linking UX features to emotion regulation outcomes, these outputs alone do not yet constitute actionable design guidance. To enable systematic translation into design decisions, the next stage focuses on integrating these findings with psychological theory and stakeholder considerations to establish a shared foundational plan.

**Stage 2: Establishing a Foundational Plan and Roadmap**

Gen-AI was prompted to simulate a theoretical integration and stakeholder map, bridging emotional regulation mechanisms with UX and technical design needs. Following the extraction of themes from the systematic literature review, the researchers leveraged Gen-AI to translate key findings into a structured foundational plan. This involved mapping user profiles, emotional needs, and aligning design goals with stakeholder expectations. The process began by prompting Gen-AI to contextualize emotion regulation (ER) within the lived experiences of individuals with ADHD exploring how they perceive, express, and manage emotions in digital contexts. The aim was to co-create a shared roadmap that articulates psychological objectives, informs design strategies, and clarifies stakeholder roles from the outset. Prompts for this stage:

- *"Based on the systematic review insights, explain how individuals with ADHD typically manage emotions, and what digital interventions can realistically support emotion regulation without increasing cognitive overload."*
- *"Using the extracted themes, formulate clear user experience and emotion regulation goals for a digital intervention. Use Dialectical Behaviour Therapy, Self-Determination Theory, and the Capability–Opportunity–Motivation–Behaviour model."*
- *"Who are the primary and secondary stakeholders in a project developing digital ER tools for ADHD? Describe their roles, motivations, and data needs."*
- *"Create a step-by-step project plan linking research goal, user needs, and stakeholder engagement following a mixed-method approach (quantitative, qualitative, AI synthesis)."*

The foundational plan developed in Stage 2 aligns empirical insights with theoretical models and stakeholder needs, providing the necessary structure for deeper synthesis. Building on this foundation, the following stage applies the UXR PoV building blocks to transform theoretical and empirical knowledge into actionable design insights and UXR Play Cards.

**Stage 3: Insight Generation and Play Card Development (Building Block Integration)**

To synthesize theoretical, empirical, and AI-generated insights into actionable design principles using the *Building Blocks* model**:** Foundation, Data Collection, Insight Generation and POV. This stage translated systematic review findings and psychological theory into ten UXR Play Cards for ADHD emotion regulation (ER) design. Gen AI was used iteratively to connect theory, data, and design application. The *Foundation* integrated DBT, SDT, and COM-B frameworks to explain emotional scaffolding, autonomy, and sustained behaviour change. In *Data Collection*, Gen-AI analysed 42 empirical studies to identify recurring patterns (e.g., attention variability, feedback sensitivity, emotional overload). During *insight generation*, the model linked these patterns with psychological theory, explaining *why* specific design mechanisms regulate emotion and engagement. Finally, *POV translation* transformed insights into UXR Play Cards, each summarizing the issue, evidence, theoretical basis, and best practice. Prompts for this stage:

- *"Map DBT, SDT, and COM-B to digital emotion regulation design for ADHD."*
- *"Cluster empirical data into themes of feedback, motivation, attention, and trust."*
- *"Generate 10 Play Card content linking theory, evidence, and design rationale."*

- *"Use the example Play Cards as a reference when creating the cards."*

While UXR Play Cards formalize design-relevant insights, their impact depends on effective communication to diverse stakeholders. Accordingly, the final methodological stage focuses on translating Play Card content into coherent, evidence-based Point-of-View narratives tailored to different roles and decision-making contexts.

**Stage 4: Crafting PoV Narratives and Stakeholder Communication**

To transform Play Card insights into coherent, evidence based PoV narratives tailored to different stakeholders. The researcher uploaded narrative template from the UXR PoV Playbook [6] and used Gen-AI to refine tone, structure, and theoretical alignment. GenAI supported translating Play Card findings into clear, role-specific narratives that integrated DBT, SDT, and COM-B principles. Prompts for this stage:

- *"Check the UXR PoV template and generate ADHD-specific PoV narratives."*
- *"Refine the PoV statement for clinical accuracy and user experience clarity."*
- *"Summarise insights from Play Cards into concise stakeholder messages."*
- *"Ensure tone consistency with neurodiversity and therapeutic language."*

Together, these four stages form an integrated, AI-augmented UXR PoV process. The following section reports the results generated at each stage, illustrating how GenAI-supported synthesis progresses from evidence extraction to actionable PoV narratives for emotion regulation–focused digital interventions.

## 3 RESULTS

This study produced a structured set of findings demonstrating how Generative AI (GenAI) can support the development of User Experience Research (UXR) Points of View (PoVs) for digital emotion regulation (ER) interventions targeting adults with ADHD. Results are organized according to the four-stage AI-powered UXR PoV framework: (1) leveraging GenAI for evidence synthesis and hypothesis development, (2) establishing a foundational stakeholder-informed roadmap, (3) translating insights into actionable UXR Play Cards, and (4) crafting stakeholder-aligned PoV narratives. Across these stages, psychological theory, systematic evidence synthesis, and UX design considerations were progressively integrated to transform research insights into design-relevant guidance.

To enhance methodological robustness, four Generative AI systems were used comparatively throughout the analysis process. Although minor differences were observed in tone, depth of explanation, and interpretive framing, the systems produced broadly convergent findings across thematic extraction, hypothesis generation, stakeholder mapping, and design translation tasks. This consistency suggests that the resulting insights are not artefacts of a single model but reflect stable patterns within the evidence base. The following sections therefore present a synthesized account of these cross-model outputs, highlighting shared insights while noting interpretive considerations where relevant.

### 3.1 Leverage GenAI and the UXR PoV Framework

In the first phase, Generative AI (ChatGPT-4o, ScholarGPT variant) was employed as a cognitive collaborator to analyse 42 peer-reviewed studies on digital ER for adults with ADHD. Through iterative prompting and researcher-led validation, ten recurring UX themes were identified and mapped to theory-informed design hypotheses using psychological models such as DBT, SDT, COM-B. The synthesized hypotheses include:

- **H1**: *Simplified emotional cues reduce cognitive overload, supporting self-regulation.*
- **H2**: *Consistent AI tone and feedback build emotional trust and predictability.*
- **H3**: *Customizable goal systems enhance user autonomy and emotional agency.*
- **H4**: *Safety features like pause modes and continuity support emotional resilience.*
- **H5**: *Neuroinclusive emotion models improve relevance for diverse users.*
- **H6**: *Mindful micro-interactions support emotional regulation through pacing.*
- **H7**: *Positive reinforcement and biofeedback sustain motivation and regulation.*
- **H8**: *Adaptive content flow aligns with attentional rhythms and emotional pacing.*
- **H9**: *Transparent AI emotional cues increase user trust and clarity.*

- **H10**: *Personalized sensory feedback sustains engagement without overload*.

Each hypothesis was co-developed through GenAI-supported synthesis and researcher interpretation, ensuring theoretical fidelity and practical relevance, and serve as the foundational knowledge artifacts for subsequent stakeholder integration and design translation in later stage.

### 3.2 Establish a Foundational Plan and Road Map

In the second stage, GenAI (ChatGPT-4o, ScholarGPT variant) was guided to synthesize stakeholder perspectives essential to designing emotion regulation tools for adults with ADHD. Drawing on literature from participatory design and neuroinclusive UX, the AI generated a structured matrix mapping five core stakeholder groups ADHD users, clinicians, UX designers, researchers, and developers to their functional roles, needs, and challenges. The stakeholder matrix outlined five key groups, each with distinct roles, needs, and design challenges. ADHD users seek emotionally safe, predictable interfaces but often face cognitive overload and volatility. Clinicians need visibility into user progress yet struggle to integrate feedback into therapy. UX designers translate psychological theory into interface principles but lack neuroinclusive design heuristics. Researchers aim to synthesize behavioural data but face limited access to consistent emotional markers. Developers prioritize feasibility and interpretability while balancing transparency with adaptive system behaviour. This mapping surfaced early design tensions and guided alignment across disciplines. This matrix clarified the interdependencies across roles and revealed systemic tensions such as aligning emotional safety with technical feasibility and theory with implementation. It also informed the co-creation strategy for subsequent design phases. The complete stakeholder framework is presented in Table 2.

| Stakeholder | Role | Need | Challenge |
|---|---|---|---|
| **ADHD Users** | End-Users | Predictable, emotionally safe digital support | Managing cognitive overload and volatility |
| **Clinicians** | Supervisors | Data visibility and progress feedback | Integrating user feedback into therapy |
| **UX Designers** | Translators | Translating theory into interface principles | Absence of neuroinclusive design heuristics |
| **Researchers** | Evaluators | Synthesizing behavioural and design data | Limited access to consistent behavioural markers |
| **Developers** | Builders | Feasibility and model interpretability | Balancing transparency and adaptivity |

*Table 2. Stakeholder Mapping within the GenAI-Augmented UXR Framework*

These insights directly informed the design of practical tools developed in the next phase, where stakeholder-informed hypotheses were operationalized into GenAI-assisted UXR Play Cards, as detailed in Section 3.3.

### 3.3 Apply GenAI-Enhanced Best Practices

In the third stage, ten GenAI-informed hypotheses were transformed into a structured, research-driven tool set: the UXR Play Cards. Developed using the ScholarGPT model (ChatGPT-4o) and validated through expert review, these cards operationalize psychological theory and empirical synthesis into accessible, actionable formats for UX researchers, designers, and interdisciplinary teams.

Each card captures a core UX challenge in designing digital ER tools for adults with ADHD. The cards contain five elements: a thematic title, an illustrative quote, a defined issue type, a best practice recommendation, and the relevant UXR skill sets needed to address it. Together, these components enable seamless integration of behavioural theory with practical interface strategies. For example:

- Card 1 focuses on reducing cognitive load through simplified emotional cues, anchored in accessibility design and cognitive psychology.
- Card 2 addresses trust and emotional predictability in AI systems by recommending consistent tone and feedback patterns.
- Card 5 highlights the need for inclusive emotion modelling by training AI on neurodivergent emotional patterns.
- Card 7 promotes emotional scaffolding through positive reinforcement and motivation tracking, rooted in DBT and behavioural design.

Other cards address themes such as adaptive pacing, emotional transparency, safety, autonomy, and sustained engagement all common challenges identified through the systematic literature review. UXR skills span from trauma-informed design and emotion-centred micro interactions to flow theory, AI ethics, and explainable AI. These cards are designed to guide ideation, support co-design, and encourage ethical, theory-aligned decision-making across development stages. Most importantly, they establish a shared design language grounded grounded in both evidence and empathy. Overall, ten GenAI-augmented UXR Play Cards were generated to facilitate systematic review, co-design, and ethically grounded development.

| Card | Title | Quote | Issue Type | Best Practice | UXR Skills |
|---|---|---|---|---|---|
| **1** | Reducing Cognitive Load in ADHD Interfaces | Design emotional experiences that calm, not overwhelm. | Cognitive Overload | Simplify emotional cues; reduce multitasking. | Accessibility Design, Cognitive Psychology |
| **2** | Emotional Predictability and Trust in AI | Consistency breeds calm and trust. | AI Unpredictability | Keep tone, visuals, and feedback consistent. | Affective Computing, AI Ethics |
| **3** | Empowering Emotional Agency | Autonomy regulates emotion give users control. | User Disempowerment | Let users customize feedback, visual intensity. | Behavioural Design, Self-Determination Theory |
| **4** | Designing for Emotional Safety | Emotionally safe spaces begin with design boundaries. | Emotional Overstimulation | Include safety modes, pause features, calm colours. | Psychological Safety, Trauma-Informed UX |
| **5** | Inclusive AI Emotion Modelling | One emotion model doesn't fit all minds. | Neurotypical Bias | Train AI to recognize neurodivergent emotional patterns. | Inclusive AI Design, Data Ethics |
| **6** | Mindful Micro-Interactions | Tiny design moments shape emotional equilibrium. | Sensory Overwhelm | Use micro-animations and rhythmic transitions. | UX Micro interactions, Emotion-Centred Design |
| **7** | Feedback as Emotional Scaffolding | Feedback isn't just data it's emotional validation. | Motivation Fatigue | Apply positive reinforcement and progress tracking. | DBT Skills Training, Motivation Design |
| **8** | Adaptive Attention Flow | Attention is dynamic design to flex, not fight it. | Sustained Attention Decline | Adaptive pacing and contextual reminders. | Flow Theory, Adaptive Interface Design |
| **9** | Emotional Transparency in AI Feedback | Users trust systems that feel honest. | AI Emotional Misalignment | Make emotional state and confidence visible. | Human-AI Interaction, Explainable AI |
| **10** | Sustaining Emotional Engagement | Engagement grows when emotion feels seen. | Emotional Disconnection | Personalize prompts based on mood and history. | Emotion Design, Digital Wellbeing |

*Table 3. GenAI-derived UXR Play Cards for ADHD emotion-regulation design*

This tool set also served as a scaffold for the next and final stage of the process: crafting stakeholder-specific Point-of-View (PoV) narratives that align research evidence, emotional design goals, and system feasibility. This integration is discussed in detail in Section 3.4.

### 3.4 Craft Compelling PoV Narratives

In the final stage, stakeholder-aligned PoV narratives were developed to operationalise psychological theory (DBT, SDT, and COM-B) into actionable design strategies for neuroinclusive ADHD emotion regulation tools. ChatGPT-4o (ScholarGPT) was used in a structured, researcher-led process to support iterative phrasing, thematic clustering, and theory-to-design translation, while interpretive control remained with the research team. For ADHD users, the narratives emphasised emotionally predictable and adaptive interfaces that reduce cognitive overload through calibrated pacing and tone modulation. For clinicians, the focus centred on therapeutic integrity, requiring explainable emotional feedback summaries and real-time progress visibility that align with clinical workflows. UX designers were supported through structured heuristics and UXR Play Cards translating behavioural theory into interface principles, addressing the absence of established neuroinclusive design standards. Researchers benefitted from AI-assisted thematic synthesis and hypothesis refinement to enhance methodological transparency, while developers were guided toward interpretable behavioural logic and ethically grounded adaptive modelling that balances transparency with personalization. Across stakeholders, four cross-cutting design pillars empathy, predictability, autonomy, and accessibility emerged as a unifying framework, providing a shared conceptual blueprint for emotionally intelligent, theory-informed, and technically feasible system design.

| Stakeholder | Core Need | Primary Challenge | PoV Narrative Focus | GenAI-Supported Contribution |
|---|---|---|---|---|
| **ADHD Users** | Emotionally safe, predictable support | Cognitive overload, volatility | Adaptive pacing, tone modulation, emotional scaffolding | Iterative phrasing of “adaptive feedback” and clustering around empathy, predictability, autonomy |
| **Clinicians** | Progress visibility, therapeutic integrity | Integrating digital feedback into therapy | Explainable dashboards and interpretable summaries | Structured synthesis of feedback models and transparency framing |
| **UX Designers** | Translate DBT–SDT–COM-B into UI | Lack of neuroinclusive heuristics | Actionable interface heuristics and Play Cards | Theory-to-interface mapping and design principal generation |
| **Researchers** | Evidence synthesis, methodological rigour | Inconsistent behavioural markers | Transparent clustering and hypothesis refinement | AI-assisted thematic grouping with human validation |
| **Developers** | Feasible, interpretable adaptive systems | Balancing transparency and personalization | Explainable behavioural logic and ethical modelling | Flow structuring, adaptive logic articulation, interpretability framing |

*Table 4. AI-Supported PoV Narrative Development within the UXR PoV Framework*

The following discussion examines the methodological, theoretical, and ethical implications of this GenAI-augmented PoV approach, reflecting on its strengths, limitations, and broader contributions to neuroinclusive UXR practice.

## 4 DISCUSSION

This study explored the integration of GenAI into UXR for the design of digital ER interventions for adults with ADHD. By aligning psychological theory including DBT, SDT, and the COM-B model with GenAI-supported synthesis tools, the project aimed to build emotionally intelligent, neuroinclusive digital systems. The findings reveal that while GenAI can serve as a powerful collaborator in research synthesis, it also introduces interpretive risks that require reflexive, bias-aware practice.

Generative AI demonstrated significant value in accelerating data-driven tasks, including coding qualitative insights, clustering engagement variables, and generating evidence-informed hypotheses. These capabilities align with recent studies showing how GenAI enhances pattern detection and hypothesis formation in information-intensive domains [15]. By rapidly synthesizing thematic insights across 42 studies, GenAI contributed to the construction of 10 theoretically grounded UXR Play Cards, each representing a testable intervention mechanism. However, despite its utility in accelerating synthesis, GenAI was limited in its capacity to interpret affective nuance or engage empathetically with neurodiverse lived experience a constraint echoed in broader literature on affect modelling and sentiment analysis [16].

These limitations became more pronounced when considering stakeholder diversity. Clinicians, designers, developers, and ADHD users each introduced distinct value systems to the design process. Clinicians emphasized therapeutic safety and emotional validity, while designers focused on accessibility and usability. Developers, by contrast, prioritized feasibility, performance. This epistemic diversity, although enriching, created interpretive tensions over how “effective emotion regulation” should be defined. Such stakeholder-induced biases are common in interdisciplinary mental health design, particularly in neurodivergent contexts, where emotional experiences defy standardization [17] [18]. To navigate these tensions, the research team employed structured used Play Cards as boundary objects to anchor decisions in both evidence and empathy.

A more systemic limitation was observed in the AI models themselves. Despite cross-platform triangulation including ChatGPT, Microsoft Copilot, Google Gemini, and NotebookLM each GenAI tool exhibited distinct biases. ChatGPT produced coherent but occasionally speculative interpretations, while Copilot's outputs were rigid and mechanistic. Gemini demonstrated creative potential but lacked psychological depth. NotebookLM frequently displayed anchoring bias, overemphasizing high-frequency sources. These findings are consistent with emerging research identifying systematic biases in large language models, including alignment biases, overconfidence in AI-related domains, and representational asymmetries favouring dominant cultural narratives [19, 20].

Trabelsi et al. (2026) show that GenAI tools can consistently recommend AI-favourable solutions and overestimate the value of AI-related interventions in decision-making scenarios [19]. Similarly, Krasanakis and Papadopoulos (2024) outline how current fairness evaluation frameworks fail to account for multi-dimensional bias, especially in systems that operate across affective, behavioural, and demographic attributes [18]. These epistemic tendencies risk marginalizing neurodivergent and cross-cultural perspectives in system design a concern that is particularly salient for emotion regulation tools intended for ADHD users, whose needs frequently diverge from normative assumptions encoded in training data [21].

In response, this study adopted a researcher-led, AI-assisted methodology. Rather than delegating interpretation entirely to GenAI, human researchers retained oversight and conducted reflexive validation of all hypotheses and insights [20]. GenAI was used to scale, not substitute, analytical reasoning a strategy aligned with recent ethical guidelines that recommend a shift from principle-based to practice-oriented AI governance [22]. The role of the researcher became one of curator, reconciling AI-generated outputs with contextual knowledge, psychological theory, and stakeholder insights. This hybrid model also echoes arguments made by Mokanderz et al. (2021), who emphasize the importance of AI auditing and post-deployment monitoring, particularly in high-risk applications such as digital mental health [22].

Despite efforts to mitigate risk, a persistent tension remained: the balance between AI automation and human interpretation. While GenAI was effective at summarizing empirical patterns and proposing conceptual links, it lacked the capacity to interpret emotional subtleties, user values, or long-term engagement dynamics especially in populations where affect is unstable or atypical. As shown by Mohammad (2020), the inability of current sentiment analysis systems to generalize beyond training domains poses a risk for the interpretive validity of AI-supported insights. Human-centred design for mental health, therefore, continues to demand affective sensitivity and theoretical grounding skills that remain the domain of human researchers [23].

The implications for UXR practice are clear. GenAI offers substantial benefits in reducing cognitive load during data synthesis and supporting hypothesis exploration at scale. However, its responsible use requires explicit bias-awareness protocols. These include cross-model comparison to detect epistemic drift, multi-source triangulation to validate findings, and the inclusion of diverse perspectives especially from neurodivergent users to ground interpretation. Without these safeguards, GenAI risks amplifying existing inequities and producing solutions that are elegant in code but exclusionary in context [24].

Beyond its domain-specific findings, this study contributes methodologically to the evolving UXR PoV framework by demonstrating how purpose-built GenAI tools can scaffold each layer of the PoV pyramid from evidence synthesis and hypothesis generation to the articulation of stakeholder-facing narratives. The integration of GenAI-supported Play Cards illustrates a practical mechanism for translating complex psychological and behavioural evidence into actionable design artefacts that remain accessible across interdisciplinary teams.

## 5 CONCLUSION AND FUTURE WORK

This research presents a GenAI-augmented UXR framework that integrates psychological models (DBT, SDT, COM-B) to design emotionally adaptive systems for neurodiverse users. By aligning AI capabilities with user-centred design methods, it reframes emotion regulation as a designable process supported through personalized, explainable feedback. Future work will focus on empirical validation, cross-diagnostic generalization, and the development of open-access toolkits for neuroinclusive design. Researchers also see potential for adapting these approaches within NHS settings, particularly in areas such as digital mental health triage, self-regulation support, and blended care models. Additionally, future iterations of this research will include systematic comparisons of generative AI outputs across different platforms (e.g., paid vs. free applications), analysing variances in quality, personalization. Team will explore how multi-prompt interactions and full-document uploads influence generative outcomes, aiming to identify optimal input strategies for generating therapeutically aligned content. Team welcome collaboration to refine this framework through real-world application and participatory research.

**6 REFERENCES**

[1] Poznyak, E., & Debbané, M. (2025). Emotion regulation beyond executive and attention difficulties: impact on daily life impairments in community adolescents. *Child and Adolescent Psychiatry and Mental Health*, 19(1), 1–12. DOI: 10.1186/s13034-025-00898-1

[2] Shaw, P., Stringaris, A., Nigg, J., & Leibenluft, E. (2014). *Emotion dysregulation in attention deficit hyperactivity disorder.* American Journal of Psychiatry, 171(3), 276–293. DOI: 10.1176/appi.ajp.2013.13070966

[3] Shetty T, Kashyap H, Mehta UM, Binu VS**.** Executive function and emotion regulation in depressive and anxiety disorders: A cross-sectional study. *Indian Journal of Psychological Medicine*. Published online June 11, 2025:02537176251340586. doi:10.1177/02537176251340586 https://www.ncbi.nlm.nih.gov/pmc/articles/PMC12162532/

[4] Storetvedt M, Svendsen E, Lydersen M, Johansen AS**.** Development of a digital mental health intervention for youths with ADHD using a person-based approach. *Front Digit Health*. 2024;6:1386892. https://www.frontiersin.org/articles/10.3389/fdgth.2024.1386892/full

[5] Bernaerts, S., Van Daele, T., Carlsen, C. K., Nielsen, S. L., Schaap, J., & Roke, Y. (2024). User involvement in digital mental health: approaches, potential and the need for guidelines. Frontiers in Digital Health, 6, Article1440660. https://doi.org/10.3389/fdgth.2024.1s440660

[5] L. van Velsen, G. Ludden, C. Grünloh, "The limitations of user- and human-centered design in an eHealth context and how to move beyond them," *J. Med. Internet Res.*, vol. 24, no. 8, e36437, 2022 doi:10.2196/37341

[6] Murray AL, Thye M, Obsuth I, Cai S, Lui M, Orr C, Saravanan A. *A narrative review to identify promising approaches for digital health interventions to support emotion regulation for adolescents with Attention-Deficit/Hyperactivity Disorder.* JMIR Ment Health. 2025 doi: 10.2196/56066

[7] M. M. Linehan, *Skills Training Manual for Treating Borderline Personality Disorder*. New York, NY, USA: Guilford Press, 1993. https://pmc.ncbi.nlm.nih.gov/articles/PMC2963469

[8] R. M. Ryan and E. L. Deci, "Self-Determination Theory and the facilitation of intrinsic motivation, social development, and well-being," *Am. Psychol.*, vol. 55, no. 1, pp. 68–78, Jan. 2000, doi:10.1037//0003-066x.55.1.68. https://pubmed.ncbi.nlm.nih.gov/11392867/

[9] S. Michie, M. M. van Stralen, and R. West, "The behaviour change wheel: a new method for characterising and designing behaviour change interventions," *Implementation Sci.*, vol. 6, no. 42, 2011, doi:10.1186/1748-5908-6-42 https://pubmed.ncbi.nlm.nih.gov/21513547/

[10] Dogan H, Barsoum RM, Giff S, Dix A, Churchill E. *Defining a UX Research Point of View (POV)*. In CHI EA 2025 - Extended Abstracts of the 2025 CHI Conference on Human Factors in Computing Systems. Association for Computing Machinery; 2025:773. https://doi.org/10.1145/3706599.3706712

[11] Endsley, M. R. (1995). Toward a theory of situation awareness in dynamic systems. *Human Factors*, 37(1), 32–64. DOI:10.1518/001872095779049499

[12] Munir, A., Aved, A., & Blasch, E. (2022). Situational Awareness: Techniques, Challenges, and Prospects. *AI*, *3*(1), 55-77. https://doi.org/10.3390/ai3010005

[13] Dogan H, Giff., S, Barsoum., R.N., and Dix., A. 2026. Developing an AI-Powered UX Research Point of View (POV). In Extended Abstracts of the 2026 CHI Conference on Human Factors in Computing Systems (CHI EA '26), April 13–17, 2026, Barcelona, Spain. ACM, New York, NY, USA, 4 pages. https://doi.org/10.1145/3772363.3778773

[14] Virginia Braun and Victoria Clarke. 2006. Using thematic analysis in psychology. *Qualitative Research in Psychology* 3, 2 (2006), 77–101. https://doi.org/10.1191/1478088706qp063oa

[15] Shneiderman B**.** Human-Centered Artificial Intelligence: Reliable, Safe & Trustworthy. *International Journal of Human–Computer Interaction*. 2022;38(1):1–12 https://doi.org/10.1080/10447318.2020.1741118

[16] Ai, Q., Zhan, J., & Liu, Y. (2025). *Foundations of GenIR*. https://arxiv.org/abs/2501.02842

[17] Klinger, J., Mateos-Garcia, J., & Stathoulopoulos, K. (2020). *A narrowing of AI research?* arXiv:2009.10385. https://arxiv.org/abs/2009.10385

[18] Krasanakis, E., & Papadopoulos, S. (2024). *Towards Standardizing AI Bias Exploration*. arXiv:2405.19022. https://arxiv.org/abs/2405.19022

[19] Trabelsi, B., Shaki, J., & Kraus, S. (2026). *Pro-AI Bias in Large Language Models*. arXiv:2601.13749. https://arxiv.org/abs/2601.13749

[20] Lin, Z. (2024). *Beyond principlism: Practical strategies for ethical AI use in research practices*. arXiv:2401.15284. https://arxiv.org/abs/2401.15284

[21] Vosough, S., et al. (2025). *Altered oscillatory brain networks in ADHD: An eLORETA study*. arXiv:2512.13539. https://arxiv.org/abs/2512.13539

[22] Mokander, J., et al. (2021). *Conformity Assessments and Post-Market Monitoring in EU AI Regulation*. arXiv:2111.05071. https://arxiv.org/abs/2111.05071

[23] Mohammad, S. (2020). *Sentiment Analysis: Automatically Detecting Valence, Emotions, and Other Affectual States from Text*. https://doi.org/10.1016/B978-0-12-821124-3.00011-9

[24] Strauss, I., et al. (2025). *Real-World Gaps in AI Governance Research*. arXiv:2505.00174. https://arxiv.org/abs/2505.00174

[25]Stephen Giff, Renée Barsoum, and Huseyin Dogan. 2024. User Experience Research: Point of View Playbook. In *Extended Abstracts of the CHI Conference on Human Factors in Computing Systems (CHI EA '24)*. ACM, New York, NY, USA. https://doi.org/10.1145/3613905.3637136